\documentclass[aps,letterpaper,10pt,prb,reprint,showpacs]{revtex4-1}
\pdfoutput=1
\usepackage{braket}
\usepackage{cancel}
\usepackage{color}
\usepackage{amsmath}
\usepackage{hyperref}
\usepackage{graphicx}
\usepackage{verbatim}
\usepackage{booktabs}
\usepackage{amsmath}

\newcommand{\be}{\begin{equation}}
\newcommand{\ee}{\end{equation}}
\newcommand{\ba}{\begin{align}}
\newcommand{\ea}{\end{align}}
\newcommand{\lb}{\label}

\begin{document}

\title{Density matrix embedding theory for interacting electron-phonon systems}
\author{Barbara Sandhoefer}
\author{Garnet Kin-Lic Chan}
\email{gkchan@princeton.edu}
\affiliation{Department of Chemistry, Frick Laboratory, Princeton University, NJ 08544, USA}
\affiliation{Princeton Center for Theoretical Science, Princeton University, NJ 08544, USA}
\date{\today}

\begin{abstract}
  We describe the extension of the density matrix embedding theory (DMET) framework to coupled interacting 
  fermion-boson systems.
  This provides a frequency-independent, entanglement embedding formalism to treat bulk fermion-boson problems.
  We illustrate the concepts within the context of the one-dimensional Hubbard-Holstein model,
  where the phonon bath states are obtained from the Schmidt decomposition of a self-consistently adjusted coherent state.
  We benchmark our results against accurate density matrix renormalization group calculations.
  \end{abstract}

\maketitle

\section{Introduction}

Density matrix embedding theory (DMET) has recently been introduced as an entanglement-based, frequency-independent quantum 
embedding method for strongly coupled degrees of freedom~\cite{Knizia2012,Knizia2013,zheng2015ground}.
Similar to the earlier dynamical mean-field theory (DMFT) and its cluster variants~\cite{Georges1996,fotso2012dynamical,maier2005quantum,PhysRevLett.87.186401}, DMET reformulates a bulk problem as 
a quantum impurity problem, where the impurity is a subset of the bulk sites. However, unlike DMFT, the impurity
embedding is not based on reproducing the lattice Green's function,
but rather the entanglement between the impurity and its environment. This time-independent formulation leads to several advantages 
over the Green's function formulations of DMFT, including a lower computational cost due to the need to describe only stationary states,
and the ability to use non-trivial bulk states, for example, with topological order, to describe the entanglement
between the impurity and environment. So far, DMET has been applied with significant success to a variety of condensed matter
problems in both model and ab-initio settings, e.g. to accurate phase diagrams of the Hubbard model on
several lattices~\cite{PhysRevB.89.165134,zheng2015ground,bulik2014density}; to frustrated spin systems~\cite{fan2015cluster}; to spectral functions~\cite{booth2015spectral}; and to
compute ab-initio binding curves of bulk materials~\cite{bulik2014electron}, to name a few examples.




Coupled fermion-boson systems provide a rich setting in which to explore new correlated quantum phases~\cite{blochreview}. A prototypical example
is a system of electrons interacting with lattice phonons. Such electron-phonon coupling is
famously the mechanism for the BCS theory of superconductivity, where Migdal-Eliashberg theory and its generalizations~\cite{migdal,eliashberg,oliveira1988density}
provide a quantitative route to transition temperatures and other
properties~\cite{scalapino1966strong, bauerME}. Within this picture
the primary effect of the electron-electron interactions is to renormalize the phonon frequency, while the phonons renormalize the electron mass. When stronger Coulomb interactions
dominate, however, as is the case in unconventional superconductors such as the cuprates and fullerenes~\cite{keimer2015quantum,takabayashi2009disorder}, the interplay between electron-electron and electron-phonon interactions is less clear. Thus, there is an important demand
for numerical methods that can correctly treat both the interacting fermionic degrees of freedom, and the bosonic degrees of freedom, on an equal footing.

In this work, we describe the extension of the DMET framework to treat coupled-fermion boson systems. 
Strong electron-electron interaction physics are often modeled using an on-site interaction, such as in the Hubbard model, and the
prototypical extension to electron-phonon coupling is provided by the Hubbard-Holstein model~\cite{holstein}.  While we will not focus here on
detailed physics of the Hubbard-Holstein model itself, reserving such a more comprehensive study for future work, we will use this model to illustrate
how to extend the DMET framework to treat electron-phonon coupling specifically, and fermion-boson systems more generally. In section~\ref{sec:theory} we describe the extension of DMET to electron-phonon problems, using the Hubbard-Holstein model as a concrete example.
In section~\ref{sec:calcs} we provide some illustrate proof-of-concept calculations on the 1D Hubbard-Holstein model, comparing
to accurate DMRG calculations. Finally, in section~\ref{sec:conclusions} we describe the possible extensions of this work such as
to coupled fermion-boson systems with interacting bosons.

\section{DMET for electron-phonon systems}
\label{sec:theory}
The Hubbard-Holstein Hamiltonian is defined as
\begin{align}
  H &= \sum_{\langle ij\rangle,\sigma}\tilde{t}_{ij}c_{i\sigma}^\dagger c_{j\sigma}+\sum_{i}Un_{i\alpha}n_{i\beta}\nonumber\\
  &+\sum_{i}\omega_0\tilde{a}^\dagger_i\tilde{a}_i+\sum_{i}gn_i(\tilde{a}^\dagger_i+\tilde{a}_i), \label{eq:hhm}
\end{align}
 where $\langle ij\rangle$ denotes nearest neighbours,  $c^\dagger_{i\sigma}, {\tilde{a}^\dagger_i}$ create spin $\sigma\in\{\alpha,\beta\}$-electrons and phonons at site $i$, respectively, 
 $U$ is the on-site electron-electron interaction,  $g$,  the electron-phonon interaction, $t$, the electron hopping and $\omega_0$, the phonon frequency.
 As is common, we will work in a phonon basis that eliminates the zero-phonon mode through the shift to {$a^\dagger_i=\tilde{a}^\dagger_i+\frac{g}{\omega_0}\langle n\rangle$}, where $\langle n \rangle=\frac{N_\text{el}}{N_\text{sites}}$ is the average electronic filling.
 After
 this change in basis, the Hamiltonian becomes{, up to a constant,}
\begin{align} 
     H &= \sum_{ij,\sigma} t_{ij} c_{i\sigma}^\dagger c_{j\sigma}+\sum_i Un_{i\alpha}n_{i\beta}+\sum_i \omega_0a^\dagger_ia_i\nonumber\\ &+\sum_i g(n_i-\langle n\rangle)(a^\dagger_i+a_i),\label{eq:hhm_shift}
\end{align}
{where $t_{ij}=\tilde{t}_{ij}-\frac{2g^2}{\omega_0}\langle n\rangle n_i\delta_{ij}$}.

In DMET for ground-states~\cite{Knizia2012,Knizia2013,zheng2015ground} the ground-state   $\Psi$ and
expectation values 
of an interacting lattice Hamiltonian such as Eq.~\eqref{eq:hhm_shift} 
 are approximated by solving for
the ground-state of two coupled model problems:
a (cluster) impurity model, and an auxiliary non-interacting lattice system. We first 
introduce these two models in a qualitative fashion.
 The impurity model with Hamiltonian $H_\text{imp}$ and ground-state $\Psi_\text{imp}$, consists of a set of $N_c$ {cluster} sites (denoted $C$) cut from the interacting lattice, coupled to $N_c$ bath sites. 
 The bath sites are defined from the auxiliary non-interacting lattice system.
The auxiliary lattice Hamiltonian $h$ yields a ground-state $\Phi$, and by
performing  a Schmidt decomposition between the cluster and the remaining (environment) sites in $\Phi$ we obtain a set of bath states. For quadratic $h$, the Hilbert space
of these bath states can be identified with the Hilbert space of a set of {\it single-particle bath sites} which become the bath sites in the impurity model.
The bath sites capture the  relevant one-particle physics (e.g. hybridization effects)  
between the bare cluster and its environment in the impurity problem. 
Finally, a self-consistency condition links $h$ and $H_\text{imp}$, where
 the parameters of $h$ are varied to match the local cluster expectation values of $\Phi$ and $\Psi_\text{imp}$. At convergence, expectation values are defined from $\Psi_\text{imp}$. 


We now illustrate the above general procedure by defining $h$ and $H_\text{imp}$ 
precisely for the Hubbard-Holstein model. 
We first  specify $h$ and the construction of the bath sites.
In  DMET for pure fermionic problems, $h$
 is typically chosen as a quadratic fermion Hamiltonian with an
 associated fermionic Gaussian (Slater determinant~\cite{Knizia2012} or BCS~\cite{zheng2015ground}) ground-state $\Phi$.
For the Hubbard-Holstein model, we choose $h$
to be the sum of quadratic electron and phonon Hamiltonians
\begin{align}
h = h_{\text{el}} + v_\text{el} + h_{\text{ph}} + v_\text{ph} + \zeta_\text{ph} \label{eq:latticeh}
\end{align}
where
\begin{align}
h_{\text{el}} &= \sum_{ij,\sigma} t_{ij} c^\dag_{i\sigma} c_{j\sigma}  \\
h_{\text{ph}} &=  \sum_i \omega_0 a^\dagger_ia_i+\sum_{\langle ij \rangle}{\varepsilon} a^\dag_i a_j-\sum_ig\langle n\rangle(a^\dagger_i+a_i) \label{eq:h_ph}\\
v_\text{el} &= \sum_C \sum_{ij \in C,\sigma} v_{\text{el},ij} c^\dag_{i\sigma} c_{j\sigma} \\
v_\text{ph} &= \sum_C \sum_{ij \in C} v_{\text{ph},ij}a^\dag_{i} a_{j} \\
\zeta_\text{ph} & = \sum_C \sum_{i \in C} \zeta_{\text{ph},i} (a^\dag_i + a_i)
\end{align}
and  $\sum_C$ denotes summation over the $N_\text{sites}/N_c$ cluster tiles (with
the ``impurity'' corresponding to $C=0$), 
and ${\varepsilon}$ is a constant to be chosen later.
The ground-state $\Phi$ of $h$ takes the form
\begin{align}
\ket{\Phi} = \ket{\Phi_\text{el}} \ket{\Phi_\text{ph}} \label{eq:latticewf}
\end{align}
where $\ket{\Phi_\text{el}}$ is the ground-state of 
$(h_{\text{el}} + v_\text{el})$ and is a 
Slater determinant as in the original DMET while $\ket{\Phi_\text{ph}}$ is the ground-state of 
$(h_{\text{ph}} + u_\text{ph} + \zeta_\text{ph})$. 

We use $\ket{\Phi_\text{el}}$ to define the electronic bath states
through its Schmidt decomposition between the impurity cluster and the remaining lattice sites (see
 the original DMET procedure~\cite{Knizia2012,Knizia2013,zheng2015ground})
\begin{align}
\ket{\Phi_\text{el}} = \sum_i \lambda_m \ket{\alpha_m}\ket{\beta_m}
\end{align}
where $\{ \ket{\alpha_m}\}$, $\{ \ket{\beta_m}\}$ denote the impurity and bath many-body 
Schmidt states. Because of the Gaussian form of $\ket{\Phi_\text{el}}$,
the impurity bath space has the special structure
\begin{align}
\{ \ket{\beta_m} \} = \mathcal{F}( \{ d_{i\sigma} \}) \otimes \prod_{j\sigma} e^\dag_{j\sigma}\ket{\mathrm{vac_\text{el}}}
\end{align}
where $\{d_{i\sigma}\}$ are a set of single-particle bath {\it orbitals} and $\mathcal{F}$ denotes
the corresponding Fock space of these orbitals~\cite{Knizia2013,zheng2015ground}. These bath orbitals together with the impurity cluster sites constitute the electron degrees of freedom in the impurity model, while, in the absence of non-local two-particle, or non-number-conserving interactions, the non-entangled environment orbitals $\{e_{i\sigma}\}$ can be ignored.
 (This is because for the $H$ under consideration, matrix elements $\langle \alpha_m \beta_m'|H|\alpha_n \beta_n'\rangle$ vanish
unless $\ket{\beta_m'}$ and $\ket{\beta_n'}$ have the same set of occupied environment orbitals, and these contribute only a constant term to the energy~\cite{zheng2015ground}).

Similarly, $\ket{\Phi_\text{ph}}$ defines phonon bath states through its Schmidt decomposition. 
However, if $\varepsilon$ in $h_{\text{ph}}$ (Eq.~\eqref{eq:h_ph}) is identically zero, then in this limit the
impurity clusters tiling the lattice have no entanglement between them,  $\ket{\Phi_\text{ph}}$ is a product state 
of the phonon vacuum on each cluster tiling the lattice: $\prod_i\ket{\text{vac}_{\text{ph},i}}$,
 and the bath states are not well defined. For infinitesimal $\varepsilon$, this degeneracy is broken, and in the absence of disorder, the phonons spread through the lattice, creating
entanglement. We choose to define $\epsilon$ to be $0_+$ in this work.  $\ket{\Phi_\text{ph}}$ is then the coherent state
\begin{align}
\vert \Phi_\text{ph}\rangle&=\exp z^\dag\ket{\text{vac}_\text{ph}} \notag\\
&=\exp(-\sum_j z_ja^\dag_j)\vert\text{vac}_\text{ph}\rangle,  \label{eq:coherent} \\
\,\text{where}\,z_j&=\sum_{ki}\zeta_kX^T_{ki}{\epsilon_i^{-1}}X_{ij}. \notag
\end{align}
$X$ and $\epsilon$ are the eigenvectors and eigenvalues from $\sum_j \left(\omega_0\delta_{ij}+\varepsilon_{ij}\right)X_{jk}=\epsilon_kX_{ik}$. 

The Schmidt decomposition of the coherent state Eq.~\eqref{eq:coherent} 
between the impurity cluster $C=0$ and its environment, can be carried out 
conveniently by dividing  {$z^\dag=\sum_{j\in C=0}z_ja_j^\dag+z_E^\dag$, where the first
  term is on the impurity cluster and the second on the remaining sites. $z_E^\dag$ defines a single bath orbital,
  which is a result of the simple structure of the coherent state representation.
  We can carry out calculations with a single phonon bath orbital, however since
  the fermionic bath consists of the same number of bath orbitals as there are impurity sites, it seems desirable
  to obtain a larger set of  $N_c$ phonon bath orbitals. To do so, we can further define an
artificial division of $z_E^\dag$ into $N_c$ components through $z_E^\dag=\sum_{j=1}^{N_c}\tilde{z}_jb_j^\dag$}
and 
\begin{align}
{b_j^\dag=\frac{\sum_{i=(j-1)N}^{jN}z_ia_i^\dag}{\sum_{i=(j-1)N}^{jN}|z_i|^2}.\label{eq:ph_operators}}
\end{align}
$N=(N_\text{sites}-N_c)/N_c$. The decomposition of Eq.~(\ref{eq:ph_operators}) to define additional phonon bath sites is not unique and others can be imagined.

The bath space is then spanned by $\{ \ket{\text{vac}_\text{ph}}, b^\dag_j \ket{\text{vac}_\text{ph}}, b^\dag_ib^\dag_j \ket{\text{vac}_\text{ph}}, \ldots \}$, where
$\{b^\dag_j \ket{\text{vac}_\text{ph}}\}$ are the phonon bath orbitals. The space of the impurity
problem is finally given by the $N_c$ electron and phonon cluster sites, and $N_c$ electron and phonon bath sites. Note that in the above Schmidt decomposition and in the coherent state 
in Eq.~\eqref{eq:coherent}
we assume no upper limit on the number of phonons. If, as in the numerical calculations below, we 
have such a cutoff, the coherent
state no longer represents the exact eigenstate of $h_\text{ph}$, and the Schmidt decomposition 
is only approximately represented by the phonon operators in Eq.~\eqref{eq:ph_operators}.

With the bath sites at hand, we can now define the impurity Hamiltonian $H_\text{imp}$. 
In DMET, there are two conventions of
how to construct the impurity Hamiltonian. The first results in an impurity Hamiltonian of the Anderson type, i.e. the many-particle interactions 
only appear on the impurity. The second results in an impurity Hamiltonian which has interactions also on the bath sites. Here we use (as  previously done
in work on lattice Hamiltonians) the first, Anderson impurity construction.
Starting from the Anderson-Holstein like lattice Hamiltonian $H'$
\begin{align}
H' = h' + \sum_{i \in {C=0}} U n_{i\alpha} n_{i\beta}+{\sum_{i\in C=0}gn_i(a_i^\dagger+a_i)}, 
\end{align}
where $h'$ is of the same form as Eq.~\eqref{eq:latticeh}, but with $v$ and $\zeta$ 
terms  restricted to sites outside the impurity, i.e. $v_\text{el}$ is replaced by
$v_\text{el}'$ where
\begin{align}
v_\text{el}'= \sum_{C\neq 0} \sum_{ij \in C,\sigma} v_{\text{el},ij} c^\dag_{i\sigma} c_{j\sigma} ,
\end{align}
and $C \neq 0$ excludes summation over the impurity cluster and similarly for $v'_\text{ph}$ and $\zeta'_\text{ph}$. Then, the impurity Hamiltonian
is given by the projection of $H'$ onto the impurity model space
\begin{align}
H_\text{imp} &= P H' P,
\end{align}
where $P$ projects onto the space of the impurity model, i.e. $P=\sum_n |\Phi(n)\rangle \langle \Phi(n)|$ where $\ket{\Phi}$ is a product state in the Fock space of the impurity and bath degrees of freedom, and $n$ is an occupancy vector. The 
 projector  effects a change of basis from the original electron and phonon basis defined by operators $\{ c^\dag_i \}$, $\{ a^\dag_i \}$, to the cluster 
plus bath operators, $\{ C^\dag_i \} = \{ c^\dag_{i \in C=0}\} \oplus \{ d^\dag_i\}$,
$\{ A^\dag_i \} = \{ a^\dag_{i \in C=0} \} \oplus \{ b^\dag_i \}$. 
After projection, $H_{\text{imp}}$ becomes
\begin{align}
H_{\text{imp}}&=\sum_{ij}T_{ij}C_{i\sigma}^\dagger C_{j\sigma}+\sum_{i\in\text{imp}}Un_{i\alpha}n_{i\beta}+\sum_{ij\in\text{bath}} V_{\text{el},ij}d_{i\sigma}^\dagger d_{j\sigma}\nonumber\\
&+\sum_{ij}\Omega_{ij}A^\dagger_i A_j+\sum_{ij\in\text{bath}}\tilde{V}_{\text{ph},ij}b^\dagger_i b_j\nonumber\\
&+\sum_{i\in\text{imp}}g(n_i-\langle n\rangle)(a^\dagger_i+a_i)+\sum_{i\in\text{bath}}{Z}_i(b^\dagger_i+b_i).\lb{eq:fci}
\end{align}
where $T$, $\Omega$, $V$ and ${Z}$ represent the matrix elements $t$, $\omega$, $v$, $\zeta$ appearing in $h'$ after projecting into the impurity plus bath site basis, and  $n_{i\sigma}=c^\dag_{i\sigma} c_{i\sigma}$. (Note $i \in \text{imp}$ is equivalent to $i \in C = 0$).

Solving the interacting impurity model defined by  $H_\text{imp}$ is much
more tractable than the interacting lattice problem. If we enforce a maximum on the 
phonon number ($\text{ph}_\text{max}$) then the eigenstate 
$H_\text{imp} \ket{\Psi_\text{imp}} = E_\text{imp} \ket{\Psi_\text{imp}}$  can be obtained 
straightforwardly by exact diagonalization. This is the 
impurity solver we use in this work, although
other solvers (such as the density matrix renormalization group and coupled cluster theory) have also 
been employed in DMET~\cite{PhysRevB.89.165134,zheng2015ground,bulik2014electron}.

The last step to specify is the DMET self-consistency, which connects $h$ and $H_\text{imp}$
and defines $v_\text{el}$, $v_\text{ph}$, and $\zeta$. These fields are fixed to best match the single-particle electron and phonon density matrices in the impurity problem corresponding
to $\ket{\Psi_\text{imp}}$ and those of the lattice wavefunction $\ket{\Phi}$.
We carry out the minimization of the Frobenius norm of the matrix
with elements
\begin{align}
{\Delta\rho^\text{el}_{ij}=\langle \Phi | C^\dag_i C_j | \Phi\rangle - \langle \Psi_\text{imp} |C^\dag_i C_j  |\Psi_\text{imp}\rangle}
\end{align}
for the electrons, and
\begin{align}
{\Delta\tilde{\rho}^\text{ph}_{ij}}&={\langle \Phi | a^\dag_i a_j | \Phi\rangle - \langle \Psi_\text{imp} |A^\dag_i A_j |\Psi_\text{imp}\rangle}\nonumber\\
&+\langle \Phi | A^\dag_i |\Phi\rangle\delta_{j(2N_c+1)} -\langle \Psi_\text{imp} | a^\dag_i |\Psi_\text{imp}\rangle\delta_{j(2N_c+1)}
\end{align}
for the phonons, where $i,j$ run over the impurity and bath sites.
The procedure is carried out self-consistently  as the new fields lead
to a new lattice Hamiltonian $h$, which leads to  new bath sites, 
a new impurity problem, and a new $H_\text{imp}$. The self-consistency can develop
multiple branches, which indicates the appearance of new phases.

As with the original fermionic DMET, the electron-phonon DMET constructed above
is exact in various limits of the Hubbard-Holstein model. First, it is exact in the limits
where the electron-phonon problem is decoupled and ordinary fermionic DMET is exact, i.e.
when $U/t \to 0, g \to 0$ (and the interacting lattice $H$ 
reduces to $h$) or $U/t \to \infty, g\to 0$ (the atomic limit). It is also exact for
$U/t \to 0, g \to \infty$ and $U/t \to \infty, g\to \infty$ as the ground-state
wavefunction reduces to the product form in Eq.~\eqref{eq:latticewf} and
the phonon part is a simple product state over localized phonon vacua. Between
these various exact limits, the DMET procedure provides a physically
motivated interpolation. 

\section{Benchmark studies}
\label{sec:calcs}
We have implemented a pilot version of the DMET electron-phonon formalism above,
using an exact diagonalization solver for the impurity problem.
To assess the above procedure numerically, we now compute a few benchmarks for the
one-dimensional Hubbard--Holstein model.
The one-dimensional Hubbard-Holstein model has been the target of extensive numerical studies. Here we will compare to 
accurate DMRG results on finite chains obtained earlier by Fehske and Jeckelman~\cite{0295-5075-84-5-57001}.
In the two limits of $g\to \infty$, $U=0$ and $g=0$, $U\to \infty$, it is analytically known that the ground-state is a Peierls insulator
and Mott insulator respectively. 
In the DMRG calculations, the two insulating phases appear with a
boundary roughly in the region of $\lambda=\frac{2g^2}{\omega_0}=U$. In addition, Fehske and Jeckelmann
observed an intermediate metallic phase which was subdivided into Luttinger liquid and
bipolaronic phases. The presence of an intermediate  phase in one-dimension is well supported 
by other numerical studies with various techniques including variable-displacement Lang-Firsov~\cite{PhysRevB.67.081102}, 
other DMRG studies~\cite{PhysRevLett.95.226401,PhysRevB.76.155114}, 
 SSE QMC~\cite{PhysRevLett.95.096401,PhysRevB.75.245103}, and variational Monte Carlo~\cite{ohgoe2014variational}. However, the order of the intermediate phase remains incompletely resolved and depends on the numerical technique used~\cite{0295-5075-84-5-57001, PhysRevB.76.155114, PhysRevLett.95.096401, PhysRevLett.95.226401, ohgoe2014variational}.
\begin{table*}[htb]
\centering
\caption{\label{tab:data} Comparison between DMET (APBC, 2-site cluster) and DMRG (OBC) energies per site for a 32-site Hubbard-Holstein chain with
  various parameters. $\text{ph}_\text{max}=8$
for all calculations. The difference between APBC and OBC boundary conditions at $U=0$, $\lambda=0$ is 1.9\%. Additionally DMET energies for 504-site Hubbard-Holstein chain (APBC, 2-site cluster) are also shown.}

\begin{tabular}{cccccc}
  \hline
  \hline
$\omega$ & $\lambda$ & $U$ & DMRG Energy & DMET Energy (32) & DMET Energy (504)  \\
\hline
0.5&1.5&0.025&-2.108&-2.136&-2.136       \\
0.5&1.1&0.025&-1.878&-1.886&-1.886       \\
0.5&0.5&0.025&-1.539&-1.546&-1.545  \\
0.5&1.5&0.4&-2.002&-2.022&-2.022       \\
0.5&1.0&0.4&-1.722&-1.732&-1.730  \\
0.5&1.4&0.8&-1.843&-1.858&-1.857  \\
0.5&0.5&1.6&-1.182&-1.188&-1.188       \\
\hline
5.0&6.4&1.0&-6.083&-6.125&-6.125      \\
5.0&4.0&1.0&-3.933&-3.945&-3.945      \\
5.0&2.0&1.0&-2.380&-2.408&-2.407 \\
5.0&4.8&2.0&-4.211&-4.229&-4.229      \\
5.0&4.0&2.0&-3.558&-3.593&-3.588 \\
5.0&6.0&4.0&-4.435&-4.473&-4.470   \\
5.0&2.0&5.0&-1.582&-1.671&-1.671      \\
\hline
\hline
\end{tabular}

\end{table*}

\begin{figure}[htb]
  \includegraphics[scale=0.4]{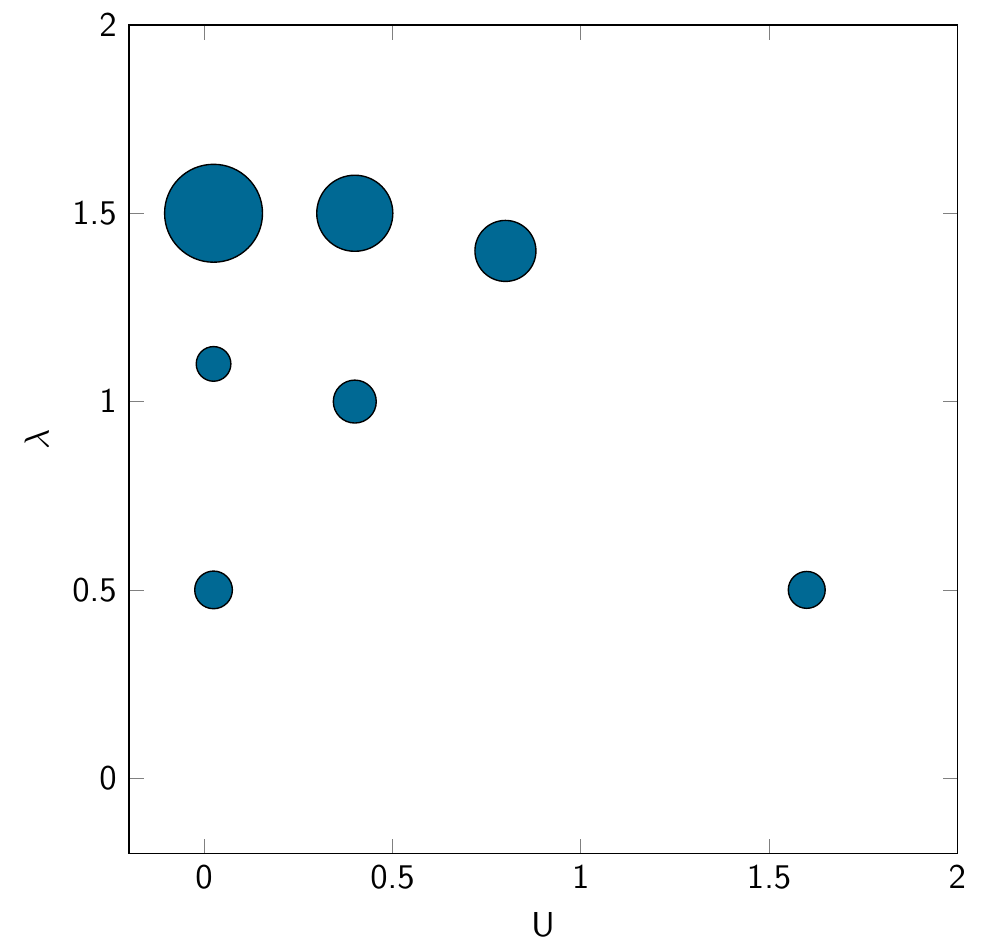}
  \includegraphics[scale=0.4]{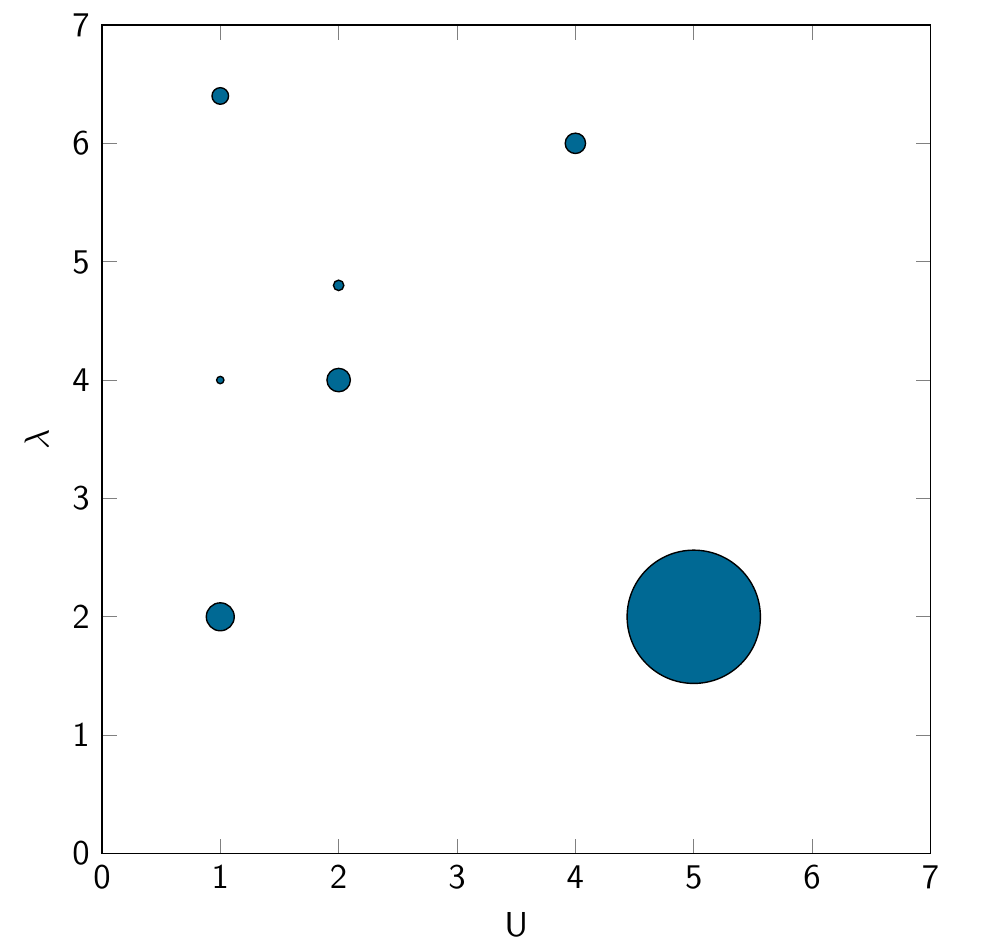}
  \caption{
Errors of the DMET calculation (32 sites, APBC) in percent of the DMRG energy (32 sites, OBC). The radius of the circle corresponds to the $0.1\times$ error. 
The left panel shows results for the adiabatic regime $\omega=0.5$, the right panel the anti-adiabatic regime $\omega=5.0$. 
For $\omega=0.5$, the largest energy error is 1.3\%, the smallest 0.46\%. For $\omega=5.0$, the largest energy error is 5.63\%, the smallest  0.31\%. 
Note that the difference in the APBC and OBC boundary conditions employed in DMET and DMRG calculations itself introduces a 1.9\% difference 
at the free fermion ($U = 0$, $\lambda=0$) level. 
\label{fig:pd}}
\end{figure}

We first compare the energies obtained from DMET with the DMRG energies of Fehske and Jeckelmann for 14 different parameters of 
a 32 site 1D Hubbard--Holstein model covering adiabatic ($\omega_0=0.5$) and anti-adiabatic ($\omega_0=5.0$) 
regimes, and various values of the coupling $\lambda=2g^2/\omega_0$ and Hubbard $U$ (all units in $t$=1). 
The results are illustrated graphically in Fig.~\ref{fig:pd} and the detailed energies per site are reported in Table~\ref{tab:data}.
The DMRG calculations used  open boundary conditions (OBC) and a phonon cutoff of $\text{ph}_\text{max}=8$ phonons per site,  while the
DMET calculations used a 2-site impurity cluster and anti-periodic boundary conditions (APBC), with the same phonon cutoff. APBC were used
to prevent an exactly zero gap for the 32-site cluster. Because
of the large phonon cutoffs, larger impurity clusters would be costly to
solve using the ED solver and are reserved for a future study.
The DMET calculations were allowed to break spin symmetry, but not number symmetry. The different boundary conditions mean that we do not expect perfect agreement; 
in the limit of free fermions ($U=0$, $\lambda=0$), the difference between OBC and APBC energies is 1.9\%. On this scale, the differences in the DMET and 
DMRG energies are  small and range between 0.5\%-5.5\% across the parameter ranges. 
DMET naturally allows a simple  extension to larger lattices than 32 sites. In Table~\ref{tab:data} we further 
show the DMET energies computed over a larger lattice of 504 sites, with the same 2-site impurity and a
maximum phonon number $\text{ph}_\text{max}=8$. 
We see that in this 1D system, the energies are in fact well converged by 32 sites, and change only by about 0.1\% going to the larger lattice, with the largest change coming in the itinerant regime. However, as seen in studies
on the 2D Hubbard model, we can expect the ability of the DMET to treat larger lattices to become important in higher dimensions, e.g. for the two-dimensional Hubbard-Holstein model~\cite{zheng2015ground,PhysRevB.89.165134,Simons2015,nowadnick2012competition}.


We now study the competition of phases at the coupling value $\lambda=4$, in the anti-adiabatic regime, $\omega = 5$.
At this coupling value, the DMRG calculations observe three phases: a Peierls insulating phase for $U < 1.5$
and intermediate phase for $1.5 < U < 3.9$, and a Mott insulating phase for $U > 3.9$. 
As mentioned above, because the DMET calculations do not break number symmetry, we can detect magnetic orders and charge orders,
but not superconducting orders.
We identify the Peierls phase as a charge-ordered phase with  order parameter
\begin{align}
\Phi_\text{co}=\frac{2}{N_c}\sum_{i=0}^{N_c/2}|n_{2i}-n_{2i+1}|\neq0
\end{align} 
and an accompanying charge-excitation gap $\Delta c_1 >0$ and the Mott phase, 
as an anti-ferromagnetic (AFM) ordered phase with order parameter
\begin{align}
\Phi_\text{afm}=\frac{1}{N_c}\sum_{i=0}^{N_c}|n^\alpha_i-n^\beta_i|\neq0,
\end{align} also
and a charge gap. 
When both orders vanish and there is no gap, we identify the phase as an intermediate phase. 
In addition to these order parameters, we can also compute a variety of other correlation functions
such as the double-occupancy on site 0, $\langle n^\alpha_0 n^\beta_0\rangle$, and the displacement $x_0 = \langle a^\dag_0 + a_0 \rangle$.
(Note that we use here the single particle gap of the auxiliary lattice system $h$ as a proxy for the charge gap. While not a rigorous measurement
it is close to the true single-particle gap in our earlier studies of the Hubbard model on the 1D, honeycomb, and square lattices~\cite{Knizia2012,booth2015spectral}). 
\begin{figure*}[htb]
  \includegraphics[width=0.23\textheight]{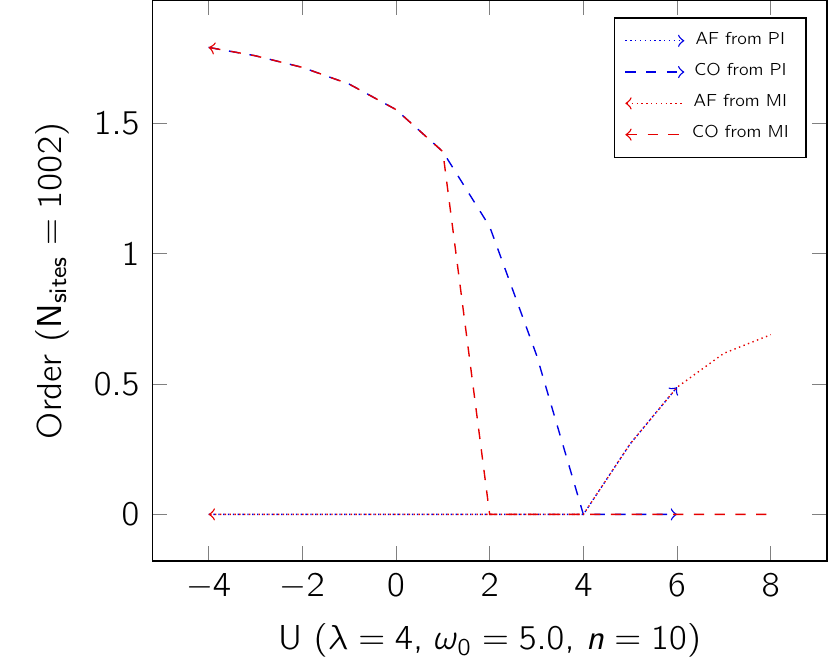}
  \hfill
  \includegraphics[width=0.225\textheight]{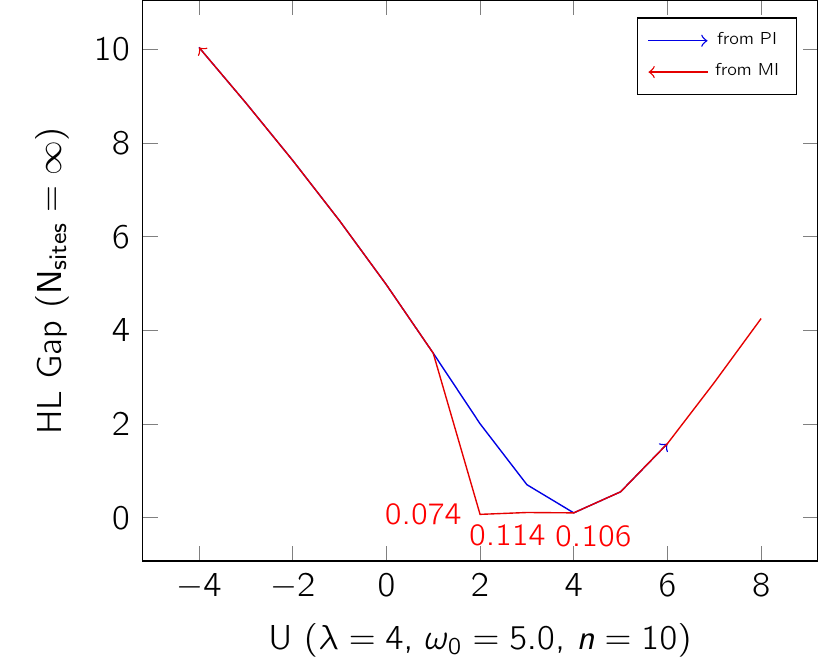}
  \hfill
  \includegraphics[width=0.245\textheight]{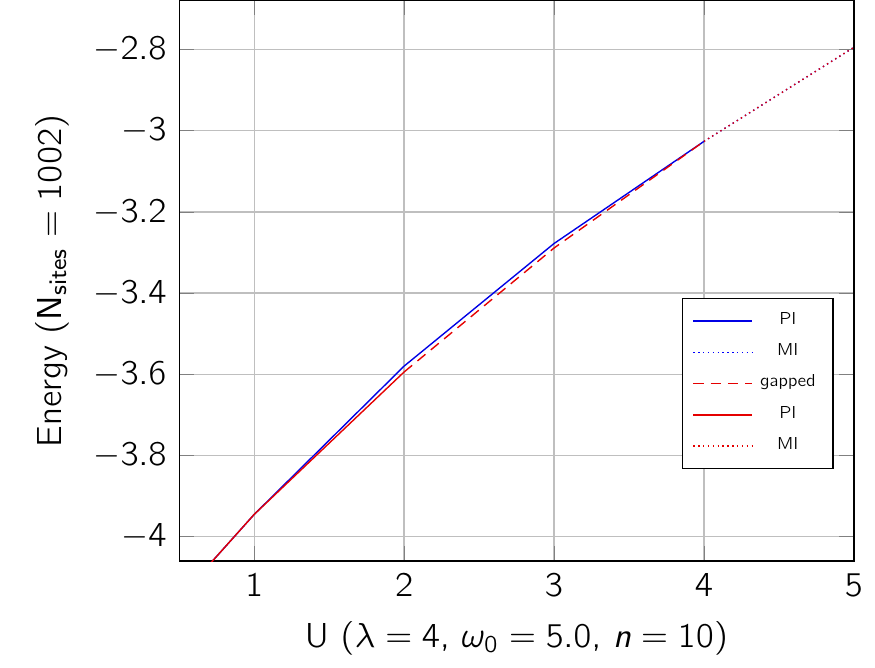}
  \caption{The left panel shows the transition from the charge-ordered to the spin-ordered phase for $\lambda=4$ and $\omega=5.0$. In the forward scan at $U=4$, the charge-order parameter vanishes, while the anti-ferromagnetic order parameter grows. The backward scan finds a (small) gapped phase without any order parameter for $U=\left[2,4\right]$. The middle panel shows the change of the HOMO-LUMO (auxiliary system single-particle) gap with $U$.
    Extrapolation from $N_\text{sites}=102, 502, 1002$ is performed to account for finite-size effects. The different phases are depicted in the right panel. The energy is converged at a maximum phonon number $\text{ph}_\text{max}=10$ and order parameters are given for the chain of length $N_\text{sites}=1002$.
    Periodic boundary conditions are used throughout. \label{fig:orderparameters_antiadiabatic}}
\end{figure*}

\begin{figure}[htb]
  \includegraphics[scale=0.48]{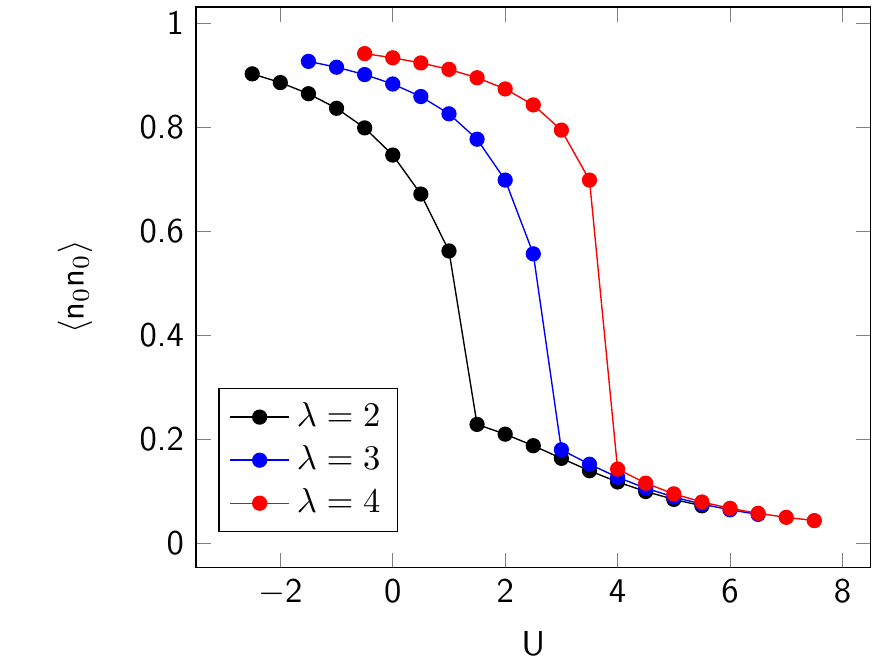}
  \includegraphics[scale=0.48]{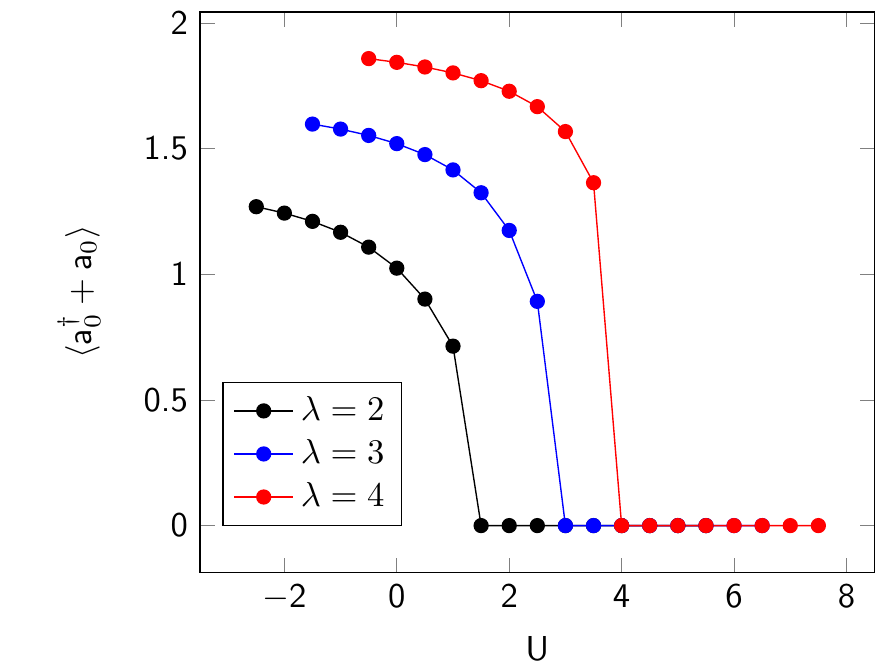}
  \caption{
  The left panel shows the double occupancy as a function of $U$ for $\lambda=2,3,4$ and $\omega_0=0.5$. The right panel shows the displacement as a function of $U$. Calculations are carried out with maximum phonon number $\text{ph}_\text{max}=10$ and chain length $\text{N}_\text{sites}=1002$ using periodic boundary conditions. \label{fig:allphases}}
\end{figure}

In the left panel of Fig.~\ref{fig:orderparameters_antiadiabatic} we show the charge-order and AFM order parameters from
a set of calculations that sweep from low U to high U, and from high U to low U. These  show clear hysteresis, indicating first
order phase transitions. The right panel shows the energies of the different DMET coexisting solutions. For $U < 2.5$, the CO (Peierls phase)
is lowest in energy. In the region $U=2.5$ to $U=4$ CO (and AFM order) is vanishing and a new phase develops.  Interestingly,
as seen from the middle panel which shows the single-particle gap, in this region the single-particle gap drops to a very small but finite
value  ($\sim 0.1t$). We identify this phase as the intermediate phase. Similar to as observed in the  DMRG studies, the intermediate phase
in the DMET calculations does not show charge or magnetic order, but has a small gap rather
than being gapless as suggested by the DMRG calculations. It is not clear whether the small gap we observe here would vanish
with larger impurity clusters.
For $U>4$ antiferromagnetic order develops, and we enter the Mott insulator phase. In Fig.~\ref{fig:allphases}
we show additional observables: the double occupancy, and the displacement. The maximum double occupancy on site $0$ is observed in the
charge ordered phase and this vanishes as $U$ increases.
Similarly we find that the displacement decreases to zero away from the CO phase. Overall, our DMET data closely corresponds to
the 3-phase picture in the DMRG calculations, and we observe similar phase boundaries.



\section{Conclusions}
\label{sec:conclusions}
In this work we described the extension of the density matrix embedding theory to systems of coupled fermions and bosons,
using the electron-phonon Hubbard-Holstein Hamiltonian as a particular example. We performed pilot calculations on the 1D Hubbard-Holstein
model using a small two-site impurity cluster, and found good agreement with the energetics of earlier benchmark DMRG calculations.
In the antiadiabatic regime, we also observed a three-phase behaviour, including an intermediate phase between the charge-ordered and Mott-insulating states,
with similar phase boundaries as found in the earlier DMRG work.

We can imagine further extensions of the ideas in this report, both respect to the physics and the methodology.
For example, here we carried out a preliminary study of the one-dimensional Hubbard-Holstein model. Extending this 
to studies in two-\cite{nowadnick2012competition} and higher-dimensions~\cite{sangiovanni2005,werner2007} is of clear interest.
Further, while the Hubbard-Holstein model only contains non-interacting phonons, the
DMET formalism is equally applicable to interacting phonons, which would
allow us to study many interesting coupled interacting fermion-boson systems, or even interacting pure boson systems,
as found, for example in cold atomic gases~\cite{blochreview}.

With respect to the DMET formulation itself, a key question to explore  is alternative definitions of
the auxiliary phonon system. While we here used a simple coherent state ground-state to define the phonon bath sites, other choices
which describe less classical phonons can be used.
Finally, on the numerical front, we  employed an exact diagonalization solver for the DMET impurity problem. Extensions
to other ground-state solvers, such as the density matrix renormalization group, or diffusion or auxiliary field quantum Monte Carlo,
would open up the possibility of more definitive calculations using larger impurities and realistic interactions.

\section{Acknowledgements}
We  are grateful to  Eric Jeckelmann and Holger Fehske for supplying the DMRG reference energies. We acknowledge helpful discussions with Boxiao Zheng.
This work was primarily supported by the US Department of Energy via DE-SC0010530. Additional support was provided by the US Department of Energy,
SciDAC DE-SC0008624, as well as through the Simons Foundation, through the Simons Collaboration on the Many-Electron Problem.
\bibliographystyle{apsrev4-1}
\bibliography{ref,literature}

\end{document}